\documentclass[twocolumn,aps,prl,showpacs,showkeys,assymb,amsmath,longbibliography]{revtex4-1}

\usepackage{color}
\usepackage{amsmath}
\usepackage{amssymb}
\usepackage{graphicx}
\usepackage[colorlinks=true,linkcolor=red]{hyperref}
\usepackage[sort&compress]{natbib}
\usepackage[dvipsnames]{xcolor}
\usepackage{soul}
\usepackage{amsfonts}
\usepackage{float}
\usepackage{color}

\begin{document}

\title{Chiral Flat Bands: Existence, Engineering and Stability}
\author{Ajith Ramachandran}
\author{Alexei Andreanov}
\author{Sergej Flach}
\affiliation{Center for Theoretical Physics of Complex Systems, Institute for Basic Science (IBS), Daejeon 34051, Republic of Korea}
\date{\today}

\begin{abstract}
We study flat bands in bipartite tight-binding networks with discrete translational invariance. Chiral flat bands with
chiral symmetry eigenenergy $E=0$ and host compact localized eigenstates for finite range hopping.
For a bipartite network with a majority 
sublattice 
chiral flat bands emerge.
We present a simple generating principle of chiral flat band networks and as a showcase add to the previously observed cases a number
of new potentially realizable chiral flat bands in various lattice dimensions. 
Chiral symmetry respecting network perturbations - including disorder and synthetic magnetic fields -  preserve both the flatband and the modified compact localized states. 
Chiral flatbands are spectrally protected by gaps, and pseudogaps in the presence of disorder due to 
Griffiths effects.
\end{abstract}

\maketitle

Hermitian tight binding translationally invariant lattices with the eigenvalue problem
$E\Psi_l = -\sum_m t_{lm}\Psi_m$  and
certain local symmetries have been shown to sustain one or a few completely dispersionless bands, called flat bands (FB) in their band structure~\cite{derzhko2015strongly}.  
Flat bands with finite range hoppings rely on the existence of a macroscopic number of degenerate compact localized eigenstates (CLS) $\{\Psi_l\}$ at the FB energy $E_{FB}$
which have strictly zero amplitudes outside a finite region of the
lattice due to destructive interference \cite{flach2014detangling}. Flat band networks have been proposed in one, two, and three dimensions and various flat band generators were
identified~\cite{mielke1991ferromagnetism,tasaki1992ferromagnetism}, which harvest on local symmetries. A recent systematic attempt to classify flat band networks through the
properties of CLS was used to obtain a systematic flat band network generator for one-dimensional two-band networks~\cite{maimaiti2017compact}. 
Experimental observations of FBs and CLS are reported in photonic waveguide networks~\cite{guzman2014experimental,vicencio2015observation,mukherjee2015observation,mukherjee2015observation1,weimann2016transport,xia2016demonstration}, exciton-polariton condensates~\cite{masumoto2012exciton,baboux2016bosonic,whittaker2017exciton},
and ultracold atomic condensates~\cite{taie2015coherent,jo2012ultracold}.  
FBs are obtained through a proper fine-tuning of the network parameters. For experimental realizations, the understanding and usage of FB protecting symmetries is therefore of high priority.

The interplay of flat bands and additional symmetries was discussed in a few publications.
Sutherland reported on a chiral flat band in the dice lattice~\cite{sutherland1986localization}.
Bergman et. al.~\cite{bergman2008band} studied topological protection of band touching which is not protected by any symmetry (i.e. accidental) in FB lattices. Green et. al.~\cite{green2010isolated}
reported on the possibility of opening a gap between a flatband and the other bands by breaking the time-reversal symmetry. 
As we will show below, gapped chiral FBs do not require broken time-reversal symmetry. 
Chiral bound states in the continuum were studied by Mur-Petit et. al.~\cite{murpetit2014chiral}. 
Poli et. al.~\cite{poli2017partial} examined the effect of breaking chiral symmetry in a two-dimensional Lieb lattice, which destroyed the flat band.
They experimentally investigated a microwave
realization of this partial chiral dimerized Lieb lattice. Leykam et. al.~\cite{leykam2017localization} studied hopping disorder - which by itself may preserve chiral symmetry -  in a one-dimensional diamond chain with 
already broken
chiral symmetry and therefore observed a finite localization length for states at the flatband energy, as opposed to strict compact localization for preserved chiral symmetry (see below). 
Recently, Read analyzed general topological winding properties of CLS 
using algebraic K-theory~\cite{read2017compactly}.

Bipartite lattices separate into two $A,B$ sublattices such that $E\Psi_l^{A,B} = -\sum_m t_{lm}\Psi_m^{B,A}$ and possess chiral symmetry (CS): if $\{\Psi^A,\Psi^B\}$ is an eigenvector to eigenenergy $E$,
then $\{\mp\Psi^A,\pm\Psi^B\}$ is an eigenvector to eigenenergy $-E$. We study chiral flat bands (CFB) with $E_{FB}=0$ in such systems, and the ways the chiral symmetry is protecting them.
Lieb theorem~\cite{lieb1989two} implies that chiral lattices with an odd number of bands always possess at least one chiral flat band, and we present a general method  to compute
the total number of CFBs. This allows us to derive a simple CFB network generating principle in various lattice dimensions. Disorder (or other perturbations) which preserve CS also preserve the CFB, and we show that
CLS survive up to modifications. CFBs are generically gapped away from other spectral parts, however, the gap is replaced by a pseudogap in case of hopping disorder due to
Griffiths effects~\cite{griffiths1969nonanalytic}.

We start with reminding of a well-known theorem on the existence of zero-energy states  for bipartite lattices~\cite{sutherland1986localization,lieb1989two}. It states that if the number $N_A$ of the majority $A$-sublattice 
sites is larger than the corresponding number $N_B$ of the 
minority $B$-sublattice, then there are at least  $\Delta N = |N_A -N_B| $ states 
$\{\Psi^A,0\}$ at energy $E=0$~\cite{sutherland1986localization,lieb1989two}, which occupy the majority sublattice only. 

Our first result concerns a translationally invariant $d$-dimensional lattice with CS, odd number $\nu$ of sites per unit cells and $1 \leq\mu_B < \mu_A < \nu$. The $\mu_A$ $A$-sites in any unit cell
are only connected with non-zero hopping terms $t_{lm}$ to the remaining $\mu_B$ $B$-sites (possibly belonging to other unit cells). 
The general band structure is given by dispersion relations $E_{\mu}(\vec{k})$ with the band index $\mu=1,...,\nu$ and $\vec{k}$
a $d$-component Bloch vector scanning the Brillouin zone.
It follows already by general CS that at least one of the bands must either cross $E=0$ (finite number of zero energy states) or be a FB at $E=0$ 
(macroscopic number of zero energy states),
since any band which does not cross $E=0$ is either positive or negative valued, and has a symmetry related partner band. Due to the odd number of bands, there is at least one \textit{unpaired}
band which therefore must transform into itself under CS action. In the following we focus on Hermitian system, but the concepts can be carried over to non-Hermitian systems as well. Further, since $\nu$ is odd, the difference in the number of sites on the $A$ and $B$ sublattices
$\Delta  N  = N_{uc} ( 2\mu_A  - \nu) \ne 0$, where $N_{uc} \sim L^d$ is the number of unit cells, and $L$ is the linear dimension.
This implies a  macroscopic degeneracy at $E=0$, which is only possible with precisely $ (2\mu_A - \nu)$ FBs at $E=0$. This observation 
suggests a natural classification of CFB by the imbalance of minority and majority sites, and
will be used for a CFB generator as we illustrate below.

\begin{figure}
    \includegraphics[clip,width=\columnwidth]{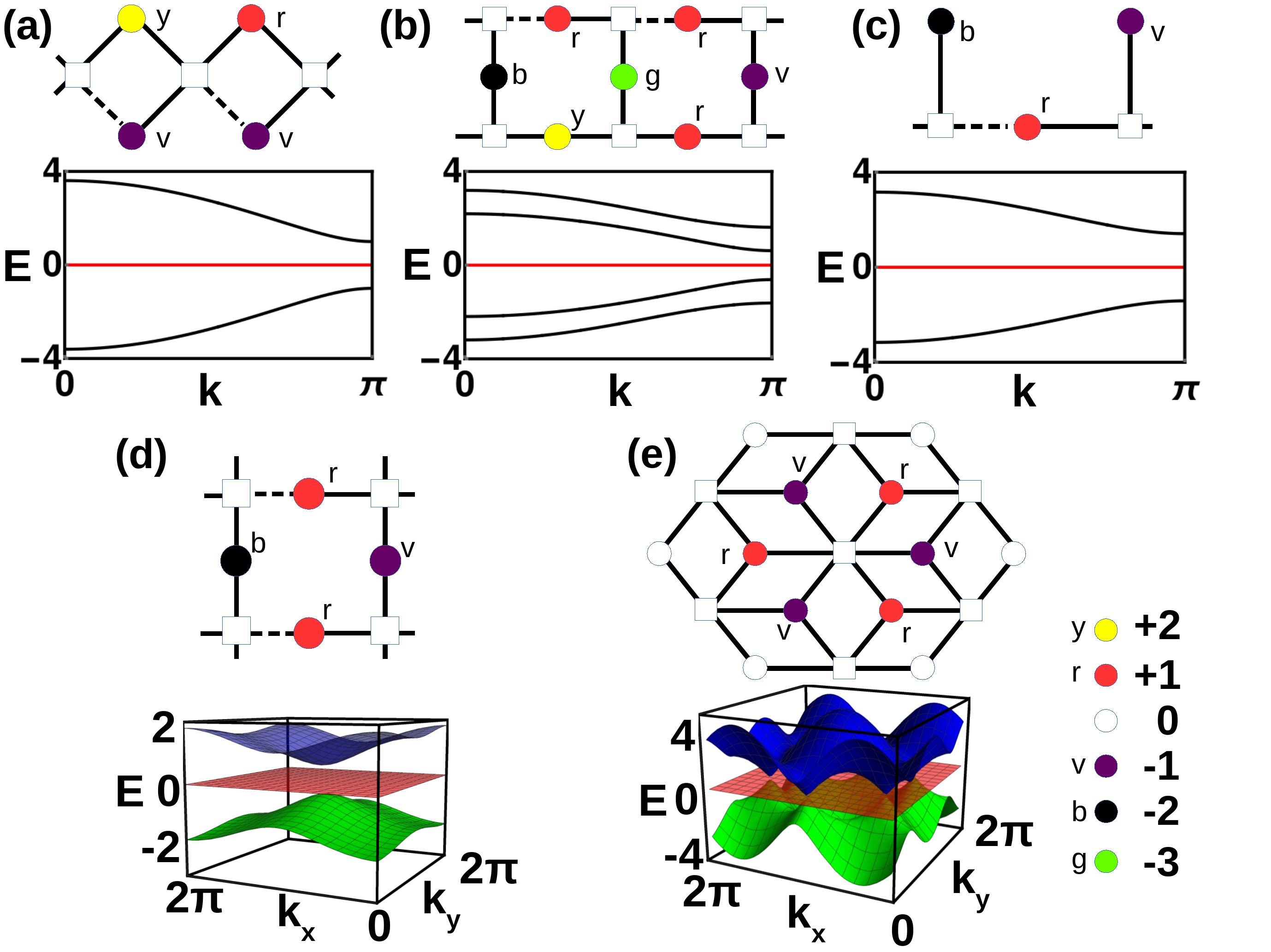}
    \caption{(Color online) Modifications of known CFB networks and their band structure (see e.g.~\cite{derzhko2015strongly,flach2014detangling}). 
Majority and minority sublattice sites are shown with 
circles and squares respectively. Solid lines: $t=1$, dashed lines: $t=2$. CLS amplitudes (not normalized) are shown in color code.
 (a) diamond, (b) 1d Lieb, (c) stub (d) 2d Lieb, (e) $\mathcal{T}_3$ (dice). 
}
    \label{fig1}
\end{figure}

\begin{figure}
    \includegraphics[clip,width=\columnwidth]{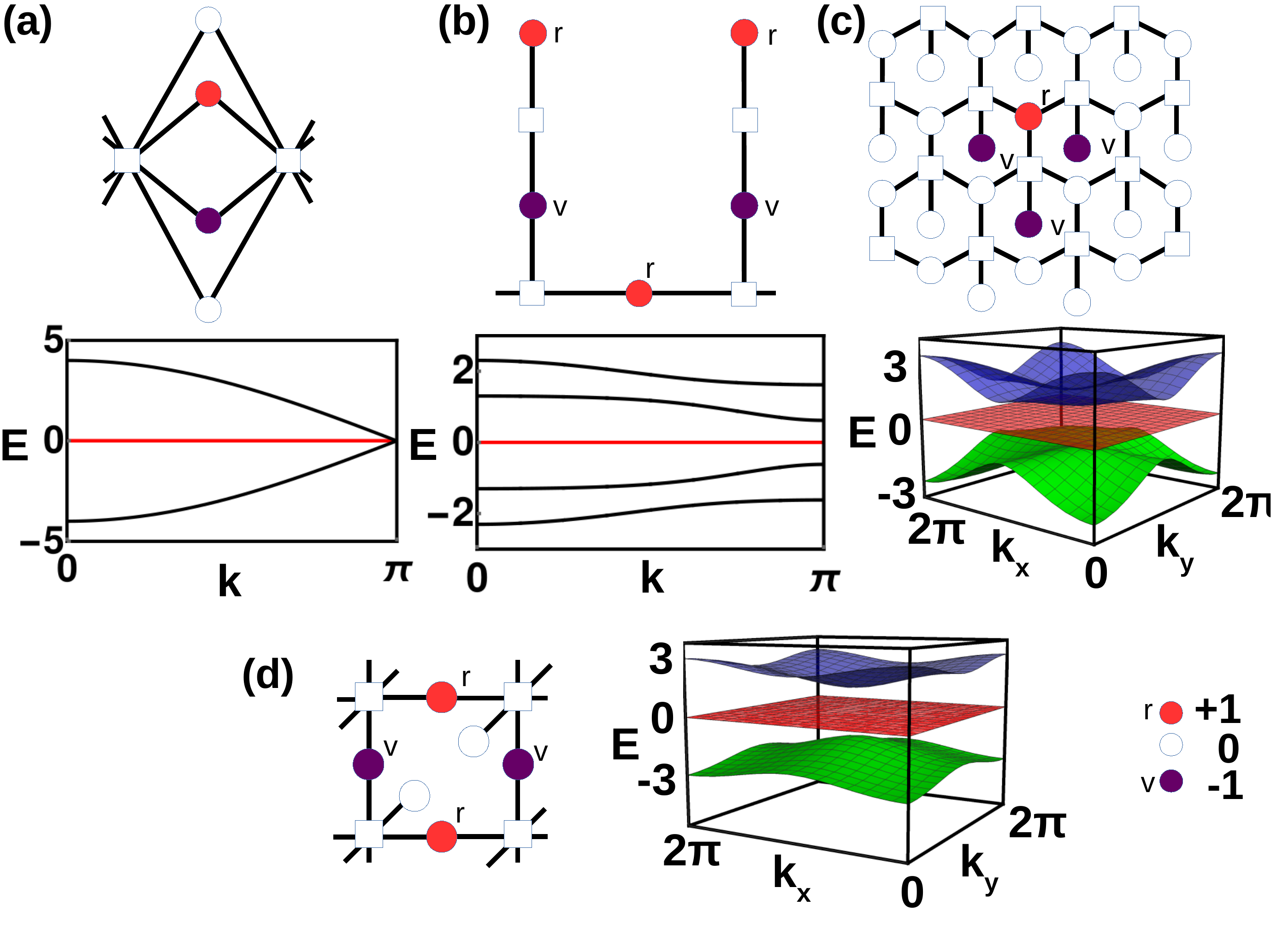}
    \caption{(Color online) Novel CFB examples (with dispersion relations). Majority and minority sublattice sites are shown with 
circles and squares respectively. CLS amplitudes (not normalized) are shown in color code.
 (a) double diamond, (b) stub3, (c) 2d stub (d) decorated Lieb
}
    \label{fig2}
\end{figure}

Let us first discuss $d=1$.
For $\nu=3$ there is only one possibility $\mu_A=2$.  A known example is the
diamond chain structure shown in Fig.~\ref{fig1}(a). The cutting of
one bond leads to the stub structure Fig.~\ref{fig1}(c). 
For $\nu=5$ there are two possibilities - $\mu_A=3,4$. The case $\mu_A=3$ leads to a
generalized Lieb structure Fig.~\ref{fig1}(b). 
Cutting a bond produces a generalized
\textit{stub3} lattice Fig.~\ref{fig2}(b). 
The second case $\mu_A=4$ arrives at a new network structure which we coin \textit{double diamond} chain Fig.~\ref{fig2}(a).

\begin{figure}
    \includegraphics[clip,width=\columnwidth]{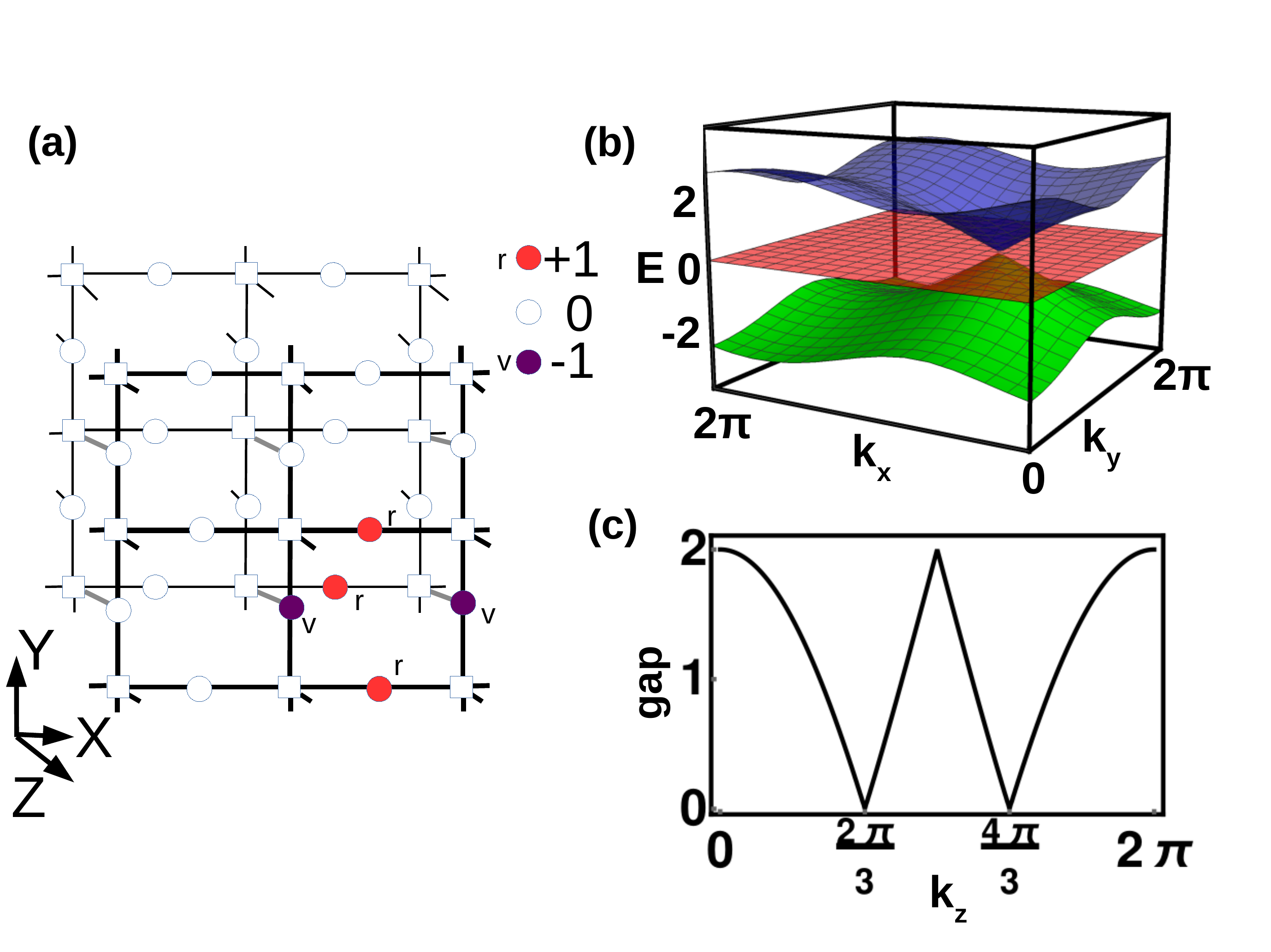}
    \caption{(Color online) A $d=3$ CFB example of the generalized 3d Lieb structure.
(with dispersion relation at fixed $k_z=2\pi/3$).
Majority and minority sublattice sites are shown with 
circles and squares respectively. CLS amplitudes (not normalized) are shown in color code.
}
    \label{fig3}
\end{figure}

For $d=2$ and $\nu=3$ case the only partitioning is again $\mu_A=2$, yet there are different choices of Bravais lattices. For the tetragonal Bravais
lattice we find a generalized 2d Lieb structure Fig.~\ref{fig1}(d). 
The hexagonal Bravais lattice yields e.g. the $\mathcal{T}_3$ or dice lattice~\cite{sutherland1986localization,vidal1998aharonov,vidal2001disorder} Fig.~\ref{fig1}(e).
Cutting two bonds in each unit cell of the $\mathcal{T}_3$ lattice yields a novel \textit{2d stub} lattice
Fig.~\ref{fig2}(c).
For $\nu=5$, there are again two partitionings $\mu_A=3,4$.  
The case $\mu_A=3$ leads to an edge centered honeycomb lattice~\cite{lan2012coexistence}. 
In the second case $\mu_A=4$ with three CFBs and two dispersive bands (decorated Lieb lattice, Fig. ~\ref{fig2}(d)).
For $d=3$ and $\nu=3$ we obtain a novel generalized \textit{3d Lieb} structure Fig.~\ref{fig3}.

\begin{figure}
    \includegraphics[clip,width=\columnwidth]{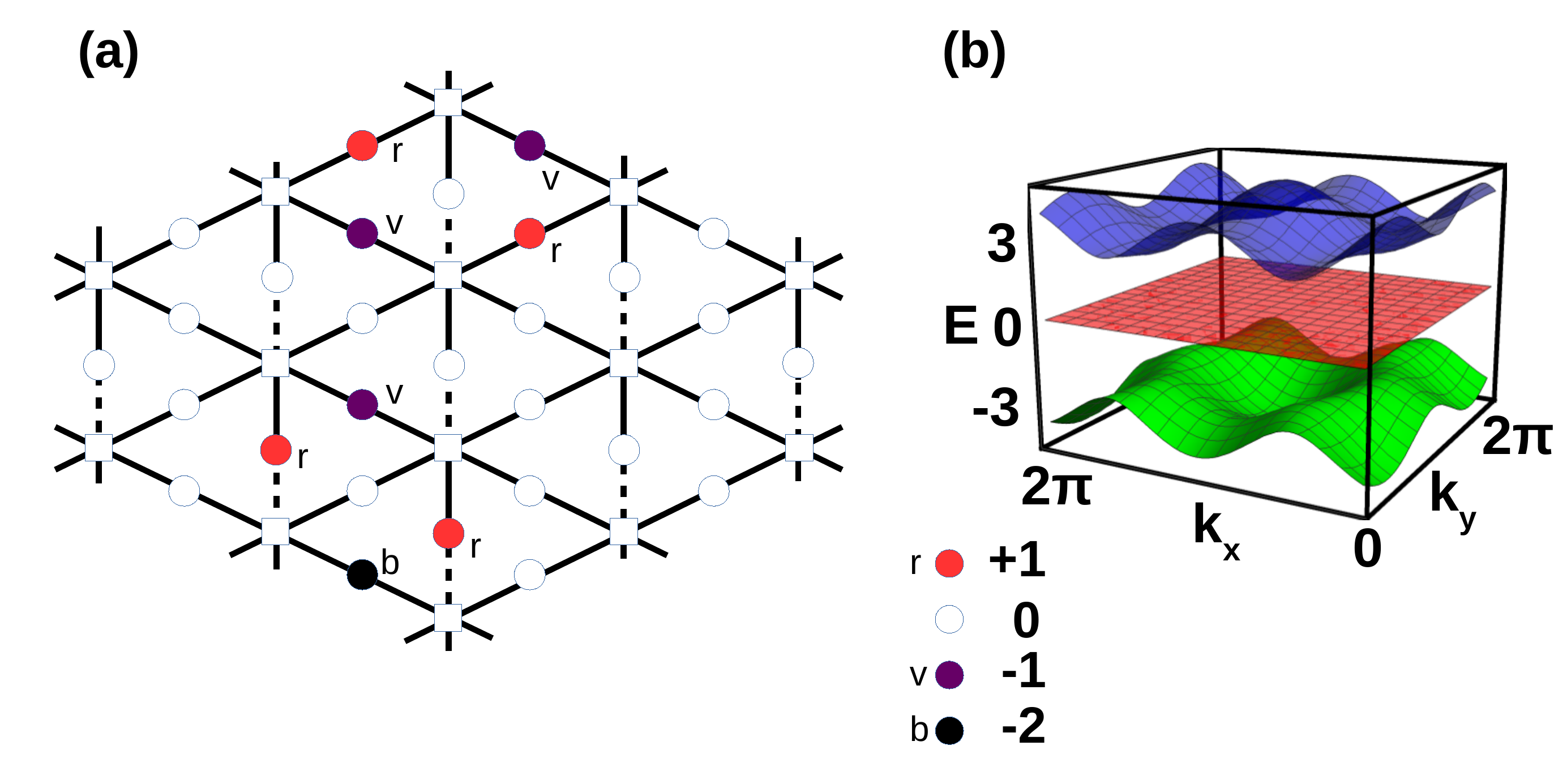}
    \caption{(Color online) A $d=2$ CFB example with four bands and three degenerated CFBs, coined $2d$ RAF lattice
(with dispersion relation). Majority and minority sublattice sites are shown with 
circles and squares respectively. CLS amplitudes (not normalized) are shown in color code.
}
    \label{fig4}
\end{figure}

The above approach can be extended to larger odd band numbers. Further,
the approach is not restricted to odd band numbers only. Any even band number $\nu \geq 4$ is working as well,
as long as $\mu_A > \mu_B$. For instance $\mu_A=3$ and $\mu_B=1$ is the first nontrivial CFB case with the smallest 
number of four bands, with two degenerated CFBs. A two-dimensional realization of such a structure is the \textit{2d RAF} (or bond-centered triangular) structure
in Fig.~\ref{fig4}.

The most general way to generate CBFs is to simply set the numbers $\nu$, $\mu_A$, $d$ and to pick a Bravais lattice.
Any values of hoppings between majority and minority sublattices are allowed. The hopping range can be anything, from compact
(or nearest neighbour) as chosen in the above examples, to exponentially decaying, or algebraic decaying, or not decaying at all in
the lattice space. For non-compact hoppings, CLS states are not expected to persist in general and turn into exponentially, algebraically or completely extended
flat band eigenstates, since the number of equations to be satisfied turns macroscopic. 
Our generator therefore provides a straightforward path of generating flat bands which lack compact localized state support.

For compact or nearest-neighbour hopping it is always possible to construct a suitably sized CLS for a CFB, as follows from a simple 
counting of equations and variables.
The nonzero CLS amplitudes must be always located on the majority sublattice and are the variables to be identified. 
The embedded and surrounding minority sites
have amplitude zero and constitute the set of equations to be satisfied.
This immediately gives an estimate of the number $n_e$ of lattice equations to be satisfied by the CLS:
$n_e = v_eL_{cls}^d + s_eL_{cls}^{d-1}$ and the number of variables $n_v = v_vL_{cls}^d$. Both numbers scale proportionally to the 
volume $L_{cls}^d$ of the CLS, with  $v_v>v_e$, and  an equation surface contribution with some proportionality factor $s_e$. For large enough volume $L_{cls}^d$ the number of equations will always be less than the number
of variables, and the CLS can be constructed. It would be interesting to combine the chiral generator with the generic CLS generator~\cite{maimaiti2017compact}.

The above examples of periodic CFB structures show that in general the CFB is gapped away from dispersive bands.
While for some cases we observe conical intersections at $E=0$ and zero gaps, this happens for highly symmetric hopping parameter sets (e.g. like all hoppings equal).
We checked, that all models discussed above show gap openings upon changing the hopping values in the corresponding sets (without destroying the network class and periodicity).
This is due to removal of accidental degeneracy of states from dispersive bands. Dispersive bands in CS networks are symmetry related,
and can touch at zero energy only for discrete sets of wavevector values. These touchings (not crossings) are additional degeneracies
which are removed by perturbations, even those that respect CS. The CFB however remains at zero energy.

Conical intersection points in two dimensional chiral networks without majority sublattices ($\mu_A=\mu_B$) are known to be protected by the very chiral symmetry.
Indeed, in this case the Hamiltonian in $k$ space is taking the form $H(k) = \left(  \begin{array}{cc}
   0 & \mathcal{T}(k) \\
   \mathcal{T}^{\dagger}(k) & 0 \\
  \end{array} 
\right)$
where $\mathcal{T}(k)$ is a square matrix of rank $\mu_A=\mu_B$. Conical intersections points in $H(K)$ are protected since the zeros
of the analytical function det $T(k)$ (and hence the zero modes of $T(k)$) survive under small perturbations of the hoppings.

However, in the case of CFBs $\mu_A \neq \mu_B$ and therefore $\mathcal{T}(k)$ is a rectangular matrix. Yet we can always represent a CFB Hamiltonian  in the form
\begin{gather}
H(k) = \left(  \begin{array}{ccc}
   \mathcal{D}_1(k) & \mathcal{Q}(k) & 0 \\
   \mathcal{Q}^{\dagger}(k) & \mathcal{D}_2(k) & 0 \\
0 & 0 & E_{FB} 
  \end{array} 
\right).
\end{gather}
In order to have a symmetry protected conical intersection point, we have to request bipartite symmetry in the subsystem of the dispersive states, i.e. $\mathcal{D}_{1,2}(k)=0$. That results in $\mu_B$ functions of $k$ which
have to vanish, with only a finite number of variables (the hopping set) at hand. Therefore the conical intersections observed for CFB are not protected by symmetry, although they might be
preserved upon perturbations along
a subset of the hopping control parameter space.

\begin{figure}
    \includegraphics[clip,width=\columnwidth]{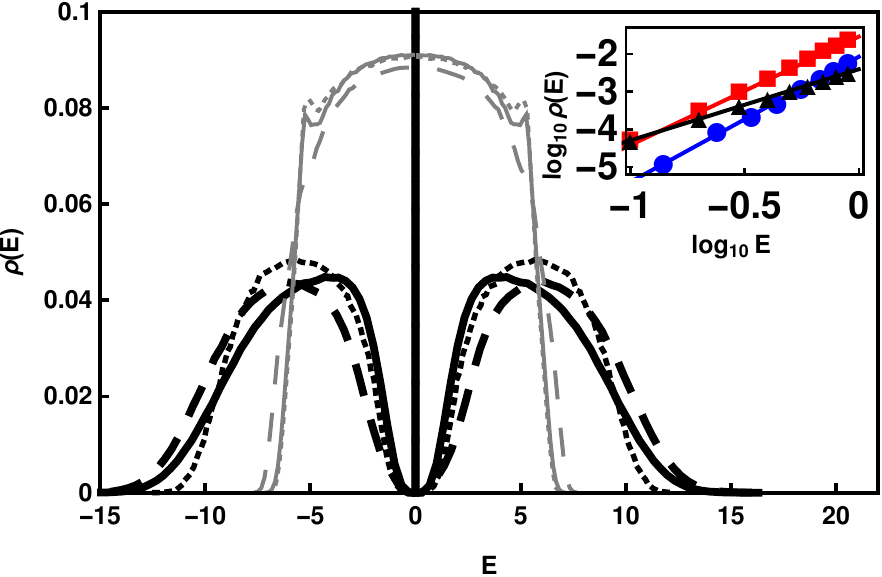}
    \caption{(Color online) The density of states $\rho(E)$ versus energy $E$ for disordered CFB networks (see text for details): 
    The thick black curves in the main plot show the density of states for the  diamond chain (solid line), 
    2d Lieb lattice (dashed line) and $T_3$ lattice (dotted line) in the presence of CS preserving hopping disorder.
The thin gray curves in the main plot show the density of states for the same systems, but in the presence of CS breaking onsite disorder. Clearly the FB is destroyed, densities turn finite,
and the Lifshitz tails are removed. The inset shows the details of the vanishing of the density of states for small but nonzero energies.
(log-log plots for diamond chain (red-squares),  2d Lieb (blue-circles) and, $\mathcal{T}_3$ (black-triangles) lattice.)
}
    \label{fig5}
\end{figure}

The CFB is protected even when destroying translational invariance while keeping CS. This can be easily
done by randomizing the hoppings, e.g. using random uncorrelated and uniformly distributed variables
$\epsilon_{ij} \in [-W/2,W/2]$ such that $t_{ij}\rightarrow t_{ij}(1+\epsilon_{ij})$. 
We first consider the 1d diamond chain from Fig.~\ref{fig1}(a) with $W=10$, which is much larger than the gap of the ordered case, and would be expected to smear out the gap completely. 
Figure~\ref{fig5} shows the density of states for this case. We clearly observe a persistent and protected CFB with $\rho(0)=\infty$ due to flatbands and pseudogap behavior $\rho(E\to 0)\to E^\alpha$ 
with model dependent exponent $\alpha$. 
The CLS persist and have a structure similar to the clean case shown on Fig.~(\ref{fig1} and~\ref{fig2}) but with amplitudes which are derived from the random hopping values $t_{ij}$ which
are connecting CLS sites with the minority sublattice. Consequently CLS are now different in different parts of the chain.
In figure ~\ref{fig5} we show the analogous results for the 2d Lieb lattice. Again the CFB is protected
by a gap, and CLS states persist, which have a structure similar to the one shown in Fig.~\ref{fig1}(d).
Finally in Fig.~\ref{fig5} we show the density of states for the disordered $\mathcal{T}_3$ lattice.
Clearly in all above cases the CFB persists symmetry preserving hopping disorder, and is protected from other states.
Notably the gap is smeared and replaced by a pseudogap due to Griffiths effects, due to rare regions that are almost translationally invariant or contain conical intersections. 
The pseudogap scaling close to $E=0$ is shown in the inset in Fig.~\ref{fig5}.

In order to demonstrate the importance of the the chiral symmetry protection, we computed densities of states for the modified eigenvalue problem - $(E-\epsilon_l)\Psi_l = -\sum_m t_{lm}\Psi_m$
- with diagonal (onsite) disorder. The random uncorrelated onsite energies $\epsilon_l$
break the CS and the CFB is destroyed, and the pseudogap and $\delta$-peak at E=0 are smeared. (Fig.~\ref{fig5}).

Finally let's discuss the possibility of linear dependence of the CLS.
To enforce linear dependence of the set of all CLS, we need to zero at least one linear combination of them, which leads to $N\mu_A$ equations with only $N$ variables (coefficients) available.
This is in general impossible, unless additional constraints are met.
Therefore the set of all CLS is generically linearly independent and spans the entire Hilbert space of the CFB. 

That is at variance with flat bands in systems lacking chiral symmetries, e.g. for the kagome and 2d pyrochlore (checkerboard) lattices~\cite{leykam2017localization}. In these cases,
the CLS set is linearly dependent~\cite{bergman2008band}. The search for one missing state leads to the existence of two different compact localized lines. The unexpected
additional state at the flat band energy is therefore due to band touching. A similar linear dependence of the CLS set happens also for non-gapped CFBs, e.g.
for the 2d Lieb lattice with all hoppings being equal, or for the dice lattice in Fig.~\ref{fig1}. Reducing the symmetry in the hopping network $\{t_{lm}\}$
while keeping the chiral symmetry preserves the flat band, opens a gap, and turns the CLS set into a linearly independent one. 
It is an interesting question whether similar reductions of the symmetry of the hopping networks for non-chiral systems with band touchings will preserve the flat band, gap it away from
dispersive states, and make the CLS set complete.

Turning the hopping matrix elements complex will destroy time reversal symmetry, and may correspond to the introduction of synthetic magnetic fields
in the context of Bose-Einstein
condensates~\cite{lin2009bose,lin2009synthetic,aidelsburger2013realization,miyake2013realizing,aidelsburger2015measuring,bloch2012quantum,struck2012tunable,celi2014synthetic}, and of light propagation
in waveguide networks~\cite{golshani2014impact,longhi2013effective}.
These changes preserve CS and therefore, the CFB persists also together with the CLS. This has been shown for the one-dimensional diamond chain in Ref.~\cite{khomeriki2016landau}, where a magnetic field
preserves the CFB, the CLS, and opens a gap. Further, we can even leave Hermitian grounds and consider dissipative couplings \cite{leykam2017flat}. Still the CFB will be protected due to the above reasoning of having a majority sublattice. Therefore CFBs can be realized even in dissipative nonHermitian settings.

To conclude, we presented the theory of chiral flat bands.
We study flat bands in chiral bipartite tight-binding networks with discrete translational invariance. Chiral flat bands are located at the
chiral symmetry eigenenergy $E=0$ and host compact localized eigenstates.
For a bipartite network with a majority 
sublattice degenerated chiral flat bands exist.
We derived a simple generating principle of chiral flat band networks and illustrated the method by adding to the previously observed cases a number
of new potentially realizable chiral flat bands in various lattice dimensions. We have also pointed out the possible constructions of flatband models with no CLS. 
Chiral symmetry respecting network perturbations - including disorder, synthetic magnetic fields, and even nonHermitian extensions -  preserve the flat band and compact localized states, which are only modified.
Chiral flat bands are thus protected and however the gaps are replaced by pseudogaps in presence of disorder, due to the contribution of rare regions.

\begin{acknowledgments}
    SF thanks Henning Schomerus for useful discussions. This work was supported by the Institute for Basic Science, Project Code (IBS-R024-D1).
\end{acknowledgments}

\bibliography{flatband}

\begin{thebibliography}{40}%
\makeatletter
\providecommand \@ifxundefined [1]{%
 \@ifx{#1\undefined}
}%
\providecommand \@ifnum [1]{%
 \ifnum #1\expandafter \@firstoftwo
 \else \expandafter \@secondoftwo
 \fi
}%
\providecommand \@ifx [1]{%
 \ifx #1\expandafter \@firstoftwo
 \else \expandafter \@secondoftwo
 \fi
}%
\providecommand \natexlab [1]{#1}%
\providecommand \enquote  [1]{``#1''}%
\providecommand \bibnamefont  [1]{#1}%
\providecommand \bibfnamefont [1]{#1}%
\providecommand \citenamefont [1]{#1}%
\providecommand \href@noop [0]{\@secondoftwo}%
\providecommand \href [0]{\begingroup \@sanitize@url \@href}%
\providecommand \@href[1]{\@@startlink{#1}\@@href}%
\providecommand \@@href[1]{\endgroup#1\@@endlink}%
\providecommand \@sanitize@url [0]{\catcode `\\12\catcode `\$12\catcode
  `\&12\catcode `\#12\catcode `\^12\catcode `\_12\catcode `\%12\relax}%
\providecommand \@@startlink[1]{}%
\providecommand \@@endlink[0]{}%
\providecommand \url  [0]{\begingroup\@sanitize@url \@url }%
\providecommand \@url [1]{\endgroup\@href {#1}{\urlprefix }}%
\providecommand \urlprefix  [0]{URL }%
\providecommand \Eprint [0]{\href }%
\providecommand \doibase [0]{http://dx.doi.org/}%
\providecommand \selectlanguage [0]{\@gobble}%
\providecommand \bibinfo  [0]{\@secondoftwo}%
\providecommand \bibfield  [0]{\@secondoftwo}%
\providecommand \translation [1]{[#1]}%
\providecommand \BibitemOpen [0]{}%
\providecommand \bibitemStop [0]{}%
\providecommand \bibitemNoStop [0]{.\EOS\space}%
\providecommand \EOS [0]{\spacefactor3000\relax}%
\providecommand \BibitemShut  [1]{\csname bibitem#1\endcsname}%
\let\auto@bib@innerbib\@empty
\bibitem [{\citenamefont {Derzhko}\ \emph {et~al.}(2015)\citenamefont
  {Derzhko}, \citenamefont {Richter},\ and\ \citenamefont
  {Maksymenko}}]{derzhko2015strongly}%
  \BibitemOpen
  \bibfield  {author} {\bibinfo {author} {\bibfnamefont {Oleg}\ \bibnamefont
  {Derzhko}}, \bibinfo {author} {\bibfnamefont {Johannes}\ \bibnamefont
  {Richter}}, \ and\ \bibinfo {author} {\bibfnamefont {Mykola}\ \bibnamefont
  {Maksymenko}},\ }\bibfield  {title} {\enquote {\bibinfo {title} {Strongly
  correlated flat-band systems: The route from heisenberg spins to hubbard
  electrons},}\ }\href {\doibase 10.1142/S0217979215300078} {\bibfield
  {journal} {\bibinfo  {journal} {Int. J. Mod. Phys. B}\ }\textbf {\bibinfo
  {volume} {29}},\ \bibinfo {pages} {1530007} (\bibinfo {year}
  {2015})}\BibitemShut {NoStop}%
\bibitem [{\citenamefont {Flach}\ \emph {et~al.}(2014)\citenamefont {Flach},
  \citenamefont {Leykam}, \citenamefont {Bodyfelt}, \citenamefont {Matthies},\
  and\ \citenamefont {Desyatnikov}}]{flach2014detangling}%
  \BibitemOpen
  \bibfield  {author} {\bibinfo {author} {\bibfnamefont {Sergej}\ \bibnamefont
  {Flach}}, \bibinfo {author} {\bibfnamefont {Daniel}\ \bibnamefont {Leykam}},
  \bibinfo {author} {\bibfnamefont {Joshua~D.}\ \bibnamefont {Bodyfelt}},
  \bibinfo {author} {\bibfnamefont {Peter}\ \bibnamefont {Matthies}}, \ and\
  \bibinfo {author} {\bibfnamefont {Anton~S.}\ \bibnamefont {Desyatnikov}},\
  }\bibfield  {title} {\enquote {\bibinfo {title} {Detangling flat bands into
  fano lattices},}\ }\href {http://stacks.iop.org/0295-5075/105/i=3/a=30001}
  {\bibfield  {journal} {\bibinfo  {journal} {EPL (Europhysics Letters)}\
  }\textbf {\bibinfo {volume} {105}},\ \bibinfo {pages} {30001} (\bibinfo
  {year} {2014})}\BibitemShut {NoStop}%
\bibitem [{\citenamefont {Mielke}(1991)}]{mielke1991ferromagnetism}%
  \BibitemOpen
  \bibfield  {author} {\bibinfo {author} {\bibfnamefont {A}~\bibnamefont
  {Mielke}},\ }\bibfield  {title} {\enquote {\bibinfo {title} {Ferromagnetism
  in the hubbard model on line graphs and further considerations},}\ }\href
  {http://stacks.iop.org/0305-4470/24/i=14/a=018} {\bibfield  {journal}
  {\bibinfo  {journal} {J. Phys. A: Math. Gen.}\ }\textbf {\bibinfo {volume}
  {24}},\ \bibinfo {pages} {3311} (\bibinfo {year} {1991})}\BibitemShut
  {NoStop}%
\bibitem [{\citenamefont {Tasaki}(1992)}]{tasaki1992ferromagnetism}%
  \BibitemOpen
  \bibfield  {author} {\bibinfo {author} {\bibfnamefont {Hal}\ \bibnamefont
  {Tasaki}},\ }\bibfield  {title} {\enquote {\bibinfo {title} {Ferromagnetism
  in the hubbard models with degenerate single-electron ground states},}\
  }\href {\doibase 10.1103/PhysRevLett.69.1608} {\bibfield  {journal} {\bibinfo
   {journal} {Phys. Rev. Lett.}\ }\textbf {\bibinfo {volume} {69}},\ \bibinfo
  {pages} {1608--1611} (\bibinfo {year} {1992})}\BibitemShut {NoStop}%
\bibitem [{\citenamefont {Maimaiti}\ \emph {et~al.}(2017)\citenamefont
  {Maimaiti}, \citenamefont {Andreanov}, \citenamefont {Park}, \citenamefont
  {Gendelman},\ and\ \citenamefont {Flach}}]{maimaiti2017compact}%
  \BibitemOpen
  \bibfield  {author} {\bibinfo {author} {\bibfnamefont {Wulayimu}\
  \bibnamefont {Maimaiti}}, \bibinfo {author} {\bibfnamefont {Alexei}\
  \bibnamefont {Andreanov}}, \bibinfo {author} {\bibfnamefont {Hee~Chul}\
  \bibnamefont {Park}}, \bibinfo {author} {\bibfnamefont {Oleg}\ \bibnamefont
  {Gendelman}}, \ and\ \bibinfo {author} {\bibfnamefont {Sergej}\ \bibnamefont
  {Flach}},\ }\bibfield  {title} {\enquote {\bibinfo {title} {Compact localized
  states and flat-band generators in one dimension},}\ }\href {\doibase
  10.1103/PhysRevB.95.115135} {\bibfield  {journal} {\bibinfo  {journal} {Phys.
  Rev. B}\ }\textbf {\bibinfo {volume} {95}},\ \bibinfo {pages} {115135}
  (\bibinfo {year} {2017})}\BibitemShut {NoStop}%
\bibitem [{\citenamefont {Guzm{\'a}n-Silva}\ \emph {et~al.}(2014)\citenamefont
  {Guzm{\'a}n-Silva}, \citenamefont {Mej{\'\i}�a-Cort{\'e}s}, \citenamefont
  {Bandres}, \citenamefont {Rechtsman}, \citenamefont {Weimann}, \citenamefont
  {Nolte}, \citenamefont {Segev}, \citenamefont {Szameit},\ and\ \citenamefont
  {Vicencio}}]{guzman2014experimental}%
  \BibitemOpen
  \bibfield  {author} {\bibinfo {author} {\bibfnamefont {D}~\bibnamefont
  {Guzm{\'a}n-Silva}}, \bibinfo {author} {\bibfnamefont {C}~\bibnamefont
  {Mej{\'\i}�a-Cort{\'e}s}}, \bibinfo {author} {\bibfnamefont {M~A}\
  \bibnamefont {Bandres}}, \bibinfo {author} {\bibfnamefont {M~C}\ \bibnamefont
  {Rechtsman}}, \bibinfo {author} {\bibfnamefont {S}~\bibnamefont {Weimann}},
  \bibinfo {author} {\bibfnamefont {S}~\bibnamefont {Nolte}}, \bibinfo {author}
  {\bibfnamefont {M}~\bibnamefont {Segev}}, \bibinfo {author} {\bibfnamefont
  {A}~\bibnamefont {Szameit}}, \ and\ \bibinfo {author} {\bibfnamefont {R~A}\
  \bibnamefont {Vicencio}},\ }\bibfield  {title} {\enquote {\bibinfo {title}
  {Experimental observation of bulk and edge transport in photonic lieb
  lattices},}\ }\href {http://stacks.iop.org/1367-2630/16/i=6/a=063061}
  {\bibfield  {journal} {\bibinfo  {journal} {New J. Phys.}\ }\textbf {\bibinfo
  {volume} {16}},\ \bibinfo {pages} {063061} (\bibinfo {year}
  {2014})}\BibitemShut {NoStop}%
\bibitem [{\citenamefont {Vicencio}\ \emph {et~al.}(2015)\citenamefont
  {Vicencio}, \citenamefont {Cantillano}, \citenamefont {Morales-Inostroza},
  \citenamefont {Real}, \citenamefont {Mej\'{\i}a-Cort\'es}, \citenamefont
  {Weimann}, \citenamefont {Szameit},\ and\ \citenamefont
  {Molina}}]{vicencio2015observation}%
  \BibitemOpen
  \bibfield  {author} {\bibinfo {author} {\bibfnamefont {Rodrigo~A.}\
  \bibnamefont {Vicencio}}, \bibinfo {author} {\bibfnamefont {Camilo}\
  \bibnamefont {Cantillano}}, \bibinfo {author} {\bibfnamefont {Luis}\
  \bibnamefont {Morales-Inostroza}}, \bibinfo {author} {\bibfnamefont
  {Basti\'an}\ \bibnamefont {Real}}, \bibinfo {author} {\bibfnamefont
  {Cristian}\ \bibnamefont {Mej\'{\i}a-Cort\'es}}, \bibinfo {author}
  {\bibfnamefont {Steffen}\ \bibnamefont {Weimann}}, \bibinfo {author}
  {\bibfnamefont {Alexander}\ \bibnamefont {Szameit}}, \ and\ \bibinfo {author}
  {\bibfnamefont {Mario~I.}\ \bibnamefont {Molina}},\ }\bibfield  {title}
  {\enquote {\bibinfo {title} {Observation of localized states in lieb photonic
  lattices},}\ }\href {\doibase 10.1103/PhysRevLett.114.245503} {\bibfield
  {journal} {\bibinfo  {journal} {Phys. Rev. Lett.}\ }\textbf {\bibinfo
  {volume} {114}},\ \bibinfo {pages} {245503} (\bibinfo {year}
  {2015})}\BibitemShut {NoStop}%
\bibitem [{\citenamefont {Mukherjee}\ \emph {et~al.}(2015)\citenamefont
  {Mukherjee}, \citenamefont {Spracklen}, \citenamefont {Choudhury},
  \citenamefont {Goldman}, \citenamefont {\"Ohberg}, \citenamefont
  {Andersson},\ and\ \citenamefont {Thomson}}]{mukherjee2015observation}%
  \BibitemOpen
  \bibfield  {author} {\bibinfo {author} {\bibfnamefont {Sebabrata}\
  \bibnamefont {Mukherjee}}, \bibinfo {author} {\bibfnamefont {Alexander}\
  \bibnamefont {Spracklen}}, \bibinfo {author} {\bibfnamefont {Debaditya}\
  \bibnamefont {Choudhury}}, \bibinfo {author} {\bibfnamefont {Nathan}\
  \bibnamefont {Goldman}}, \bibinfo {author} {\bibfnamefont {Patrik}\
  \bibnamefont {\"Ohberg}}, \bibinfo {author} {\bibfnamefont {Erika}\
  \bibnamefont {Andersson}}, \ and\ \bibinfo {author} {\bibfnamefont
  {Robert~R.}\ \bibnamefont {Thomson}},\ }\bibfield  {title} {\enquote
  {\bibinfo {title} {Observation of a localized flat-band state in a photonic
  lieb lattice},}\ }\href {\doibase 10.1103/PhysRevLett.114.245504} {\bibfield
  {journal} {\bibinfo  {journal} {Phys. Rev. Lett.}\ }\textbf {\bibinfo
  {volume} {114}},\ \bibinfo {pages} {245504} (\bibinfo {year}
  {2015})}\BibitemShut {NoStop}%
\bibitem [{\citenamefont {Mukherjee}\ and\ \citenamefont
  {Thomson}(2015)}]{mukherjee2015observation1}%
  \BibitemOpen
  \bibfield  {author} {\bibinfo {author} {\bibfnamefont {Sebabrata}\
  \bibnamefont {Mukherjee}}\ and\ \bibinfo {author} {\bibfnamefont {Robert~R.}\
  \bibnamefont {Thomson}},\ }\bibfield  {title} {\enquote {\bibinfo {title}
  {Observation of localized flat-band modes in a quasi-one-dimensional photonic
  rhombic lattice},}\ }\href {\doibase 10.1364/OL.40.005443} {\bibfield
  {journal} {\bibinfo  {journal} {Opt. Lett.}\ }\textbf {\bibinfo {volume}
  {40}},\ \bibinfo {pages} {5443--5446} (\bibinfo {year} {2015})}\BibitemShut
  {NoStop}%
\bibitem [{\citenamefont {Weimann}\ \emph {et~al.}(2016)\citenamefont
  {Weimann}, \citenamefont {Morales-Inostroza}, \citenamefont {Real},
  \citenamefont {Cantillano}, \citenamefont {Szameit},\ and\ \citenamefont
  {Vicencio}}]{weimann2016transport}%
  \BibitemOpen
  \bibfield  {author} {\bibinfo {author} {\bibfnamefont {Steffen}\ \bibnamefont
  {Weimann}}, \bibinfo {author} {\bibfnamefont {Luis}\ \bibnamefont
  {Morales-Inostroza}}, \bibinfo {author} {\bibfnamefont {Basti\'{a}n}\
  \bibnamefont {Real}}, \bibinfo {author} {\bibfnamefont {Camilo}\ \bibnamefont
  {Cantillano}}, \bibinfo {author} {\bibfnamefont {Alexander}\ \bibnamefont
  {Szameit}}, \ and\ \bibinfo {author} {\bibfnamefont {Rodrigo~A.}\
  \bibnamefont {Vicencio}},\ }\bibfield  {title} {\enquote {\bibinfo {title}
  {Transport in sawtooth photonic lattices},}\ }\href {\doibase
  10.1364/OL.41.002414} {\bibfield  {journal} {\bibinfo  {journal} {Opt.
  Lett.}\ }\textbf {\bibinfo {volume} {41}},\ \bibinfo {pages} {2414--2417}
  (\bibinfo {year} {2016})}\BibitemShut {NoStop}%
\bibitem [{\citenamefont {Xia}\ \emph {et~al.}(2016)\citenamefont {Xia},
  \citenamefont {Hu}, \citenamefont {Song}, \citenamefont {Zong}, \citenamefont
  {Tang},\ and\ \citenamefont {Chen}}]{xia2016demonstration}%
  \BibitemOpen
  \bibfield  {author} {\bibinfo {author} {\bibfnamefont {Shiqiang}\
  \bibnamefont {Xia}}, \bibinfo {author} {\bibfnamefont {Yi}~\bibnamefont
  {Hu}}, \bibinfo {author} {\bibfnamefont {Daohong}\ \bibnamefont {Song}},
  \bibinfo {author} {\bibfnamefont {Yuanyuan}\ \bibnamefont {Zong}}, \bibinfo
  {author} {\bibfnamefont {Liqin}\ \bibnamefont {Tang}}, \ and\ \bibinfo
  {author} {\bibfnamefont {Zhigang}\ \bibnamefont {Chen}},\ }\bibfield  {title}
  {\enquote {\bibinfo {title} {Demonstration of flat-band image transmission in
  optically induced lieb photonic lattices},}\ }\href {\doibase
  10.1364/OL.41.001435} {\bibfield  {journal} {\bibinfo  {journal} {Opt.
  Lett.}\ }\textbf {\bibinfo {volume} {41}},\ \bibinfo {pages} {1435--1438}
  (\bibinfo {year} {2016})}\BibitemShut {NoStop}%
\bibitem [{\citenamefont {Masumoto}\ \emph {et~al.}(2012)\citenamefont
  {Masumoto}, \citenamefont {Kim}, \citenamefont {Byrnes}, \citenamefont
  {Kusudo}, \citenamefont {L{\"o}ffler}, \citenamefont {H{\"o}fling},
  \citenamefont {Forchel},\ and\ \citenamefont
  {Yamamoto}}]{masumoto2012exciton}%
  \BibitemOpen
  \bibfield  {author} {\bibinfo {author} {\bibfnamefont {Naoyuki}\ \bibnamefont
  {Masumoto}}, \bibinfo {author} {\bibfnamefont {Na~Young}\ \bibnamefont
  {Kim}}, \bibinfo {author} {\bibfnamefont {Tim}\ \bibnamefont {Byrnes}},
  \bibinfo {author} {\bibfnamefont {Kenichiro}\ \bibnamefont {Kusudo}},
  \bibinfo {author} {\bibfnamefont {Andreas}\ \bibnamefont {L{\"o}ffler}},
  \bibinfo {author} {\bibfnamefont {Sven}\ \bibnamefont {H{\"o}fling}},
  \bibinfo {author} {\bibfnamefont {Alfred}\ \bibnamefont {Forchel}}, \ and\
  \bibinfo {author} {\bibfnamefont {Yoshihisa}\ \bibnamefont {Yamamoto}},\
  }\bibfield  {title} {\enquote {\bibinfo {title} {Exciton--polariton
  condensates with flat bands in a two-dimensional kagome lattice},}\ }\href
  {http://stacks.iop.org/1367-2630/14/i=6/a=065002} {\bibfield  {journal}
  {\bibinfo  {journal} {New J. Phys.}\ }\textbf {\bibinfo {volume} {14}},\
  \bibinfo {pages} {065002} (\bibinfo {year} {2012})}\BibitemShut {NoStop}%
\bibitem [{\citenamefont {Baboux}\ \emph {et~al.}(2016)\citenamefont {Baboux},
  \citenamefont {Ge}, \citenamefont {Jacqmin}, \citenamefont {Biondi},
  \citenamefont {Galopin}, \citenamefont {Lema\^{\i}tre}, \citenamefont
  {Le~Gratiet}, \citenamefont {Sagnes}, \citenamefont {Schmidt}, \citenamefont
  {T\"ureci}, \citenamefont {Amo},\ and\ \citenamefont
  {Bloch}}]{baboux2016bosonic}%
  \BibitemOpen
  \bibfield  {author} {\bibinfo {author} {\bibfnamefont {F.}~\bibnamefont
  {Baboux}}, \bibinfo {author} {\bibfnamefont {L.}~\bibnamefont {Ge}}, \bibinfo
  {author} {\bibfnamefont {T.}~\bibnamefont {Jacqmin}}, \bibinfo {author}
  {\bibfnamefont {M.}~\bibnamefont {Biondi}}, \bibinfo {author} {\bibfnamefont
  {E.}~\bibnamefont {Galopin}}, \bibinfo {author} {\bibfnamefont
  {A.}~\bibnamefont {Lema\^{\i}tre}}, \bibinfo {author} {\bibfnamefont
  {L.}~\bibnamefont {Le~Gratiet}}, \bibinfo {author} {\bibfnamefont
  {I.}~\bibnamefont {Sagnes}}, \bibinfo {author} {\bibfnamefont
  {S.}~\bibnamefont {Schmidt}}, \bibinfo {author} {\bibfnamefont {H.~E.}\
  \bibnamefont {T\"ureci}}, \bibinfo {author} {\bibfnamefont {A.}~\bibnamefont
  {Amo}}, \ and\ \bibinfo {author} {\bibfnamefont {J.}~\bibnamefont {Bloch}},\
  }\bibfield  {title} {\enquote {\bibinfo {title} {Bosonic condensation and
  disorder-induced localization in a flat band},}\ }\href {\doibase
  10.1103/PhysRevLett.116.066402} {\bibfield  {journal} {\bibinfo  {journal}
  {Phys. Rev. Lett.}\ }\textbf {\bibinfo {volume} {116}},\ \bibinfo {pages}
  {066402} (\bibinfo {year} {2016})}\BibitemShut {NoStop}%
\bibitem [{\citenamefont {Whittaker}\ \emph {et~al.}(2017)\citenamefont
  {Whittaker}, \citenamefont {Cancellieri}, \citenamefont {Walker},
  \citenamefont {Gulevich}, \citenamefont {Schomerus}, \citenamefont
  {Vaitiekus}, \citenamefont {Royall}, \citenamefont {Whittaker}, \citenamefont
  {Clarke}, \citenamefont {Iorsh}, \citenamefont {Shelykh}, \citenamefont
  {Skolnick},\ and\ \citenamefont {Krizhanovskii}}]{whittaker2017exciton}%
  \BibitemOpen
  \bibfield  {author} {\bibinfo {author} {\bibfnamefont {C.~E.}\ \bibnamefont
  {Whittaker}}, \bibinfo {author} {\bibfnamefont {E.}~\bibnamefont
  {Cancellieri}}, \bibinfo {author} {\bibfnamefont {P.~M.}\ \bibnamefont
  {Walker}}, \bibinfo {author} {\bibfnamefont {D.~R.}\ \bibnamefont
  {Gulevich}}, \bibinfo {author} {\bibfnamefont {H.}~\bibnamefont {Schomerus}},
  \bibinfo {author} {\bibfnamefont {D.}~\bibnamefont {Vaitiekus}}, \bibinfo
  {author} {\bibfnamefont {B.}~\bibnamefont {Royall}}, \bibinfo {author}
  {\bibfnamefont {D.~M.}\ \bibnamefont {Whittaker}}, \bibinfo {author}
  {\bibfnamefont {E.}~\bibnamefont {Clarke}}, \bibinfo {author} {\bibfnamefont
  {I.~V.}\ \bibnamefont {Iorsh}}, \bibinfo {author} {\bibfnamefont {I.~A.}\
  \bibnamefont {Shelykh}}, \bibinfo {author} {\bibfnamefont {M.~S.}\
  \bibnamefont {Skolnick}}, \ and\ \bibinfo {author} {\bibfnamefont {D.~N.}\
  \bibnamefont {Krizhanovskii}},\ }\href {https://arxiv.org/abs/1705.03006}
  {\enquote {\bibinfo {title} {Exciton-polaritons in a two-dimensional lieb
  lattice with spin-orbit coupling},}\ } (\bibinfo {year} {2017}),\ \Eprint
  {http://arxiv.org/abs/1705.03006} {arXiv:1705.03006 [cond-mat.mes-hall]}
  \BibitemShut {NoStop}%
\bibitem [{\citenamefont {Taie}\ \emph {et~al.}(2015)\citenamefont {Taie},
  \citenamefont {Ozawa}, \citenamefont {Ichinose}, \citenamefont {Nishio},
  \citenamefont {Nakajima},\ and\ \citenamefont
  {Takahashi}}]{taie2015coherent}%
  \BibitemOpen
  \bibfield  {author} {\bibinfo {author} {\bibfnamefont {Shintaro}\
  \bibnamefont {Taie}}, \bibinfo {author} {\bibfnamefont {Hideki}\ \bibnamefont
  {Ozawa}}, \bibinfo {author} {\bibfnamefont {Tomohiro}\ \bibnamefont
  {Ichinose}}, \bibinfo {author} {\bibfnamefont {Takuei}\ \bibnamefont
  {Nishio}}, \bibinfo {author} {\bibfnamefont {Shuta}\ \bibnamefont
  {Nakajima}}, \ and\ \bibinfo {author} {\bibfnamefont {Yoshiro}\ \bibnamefont
  {Takahashi}},\ }\bibfield  {title} {\enquote {\bibinfo {title} {Coherent
  driving and freezing of bosonic matter wave in an optical lieb lattice},}\
  }\href {\doibase 10.1126/sciadv.1500854} {\bibfield  {journal} {\bibinfo
  {journal} {Sci. Adv.}\ }\textbf {\bibinfo {volume} {1}} (\bibinfo {year}
  {2015}),\ 10.1126/sciadv.1500854}\BibitemShut {NoStop}%
\bibitem [{\citenamefont {Jo}\ \emph {et~al.}(2012)\citenamefont {Jo},
  \citenamefont {Guzman}, \citenamefont {Thomas}, \citenamefont {Hosur},
  \citenamefont {Vishwanath},\ and\ \citenamefont
  {Stamper-Kurn}}]{jo2012ultracold}%
  \BibitemOpen
  \bibfield  {author} {\bibinfo {author} {\bibfnamefont {Gyu-Boong}\
  \bibnamefont {Jo}}, \bibinfo {author} {\bibfnamefont {Jennie}\ \bibnamefont
  {Guzman}}, \bibinfo {author} {\bibfnamefont {Claire~K.}\ \bibnamefont
  {Thomas}}, \bibinfo {author} {\bibfnamefont {Pavan}\ \bibnamefont {Hosur}},
  \bibinfo {author} {\bibfnamefont {Ashvin}\ \bibnamefont {Vishwanath}}, \ and\
  \bibinfo {author} {\bibfnamefont {Dan~M.}\ \bibnamefont {Stamper-Kurn}},\
  }\bibfield  {title} {\enquote {\bibinfo {title} {Ultracold atoms in a tunable
  optical kagome lattice},}\ }\href {\doibase 10.1103/PhysRevLett.108.045305}
  {\bibfield  {journal} {\bibinfo  {journal} {Phys. Rev. Lett.}\ }\textbf
  {\bibinfo {volume} {108}},\ \bibinfo {pages} {045305} (\bibinfo {year}
  {2012})}\BibitemShut {NoStop}%
\bibitem [{\citenamefont {Sutherland}(1986)}]{sutherland1986localization}%
  \BibitemOpen
  \bibfield  {author} {\bibinfo {author} {\bibfnamefont {Bill}\ \bibnamefont
  {Sutherland}},\ }\bibfield  {title} {\enquote {\bibinfo {title} {Localization
  of electronic wave functions due to local topology},}\ }\href {\doibase
  10.1103/PhysRevB.34.5208} {\bibfield  {journal} {\bibinfo  {journal} {Phys.
  Rev. B}\ }\textbf {\bibinfo {volume} {34}},\ \bibinfo {pages} {5208--5211}
  (\bibinfo {year} {1986})}\BibitemShut {NoStop}%
\bibitem [{\citenamefont {Bergman}\ \emph {et~al.}(2008)\citenamefont
  {Bergman}, \citenamefont {Wu},\ and\ \citenamefont
  {Balents}}]{bergman2008band}%
  \BibitemOpen
  \bibfield  {author} {\bibinfo {author} {\bibfnamefont {Doron~L.}\
  \bibnamefont {Bergman}}, \bibinfo {author} {\bibfnamefont {Congjun}\
  \bibnamefont {Wu}}, \ and\ \bibinfo {author} {\bibfnamefont {Leon}\
  \bibnamefont {Balents}},\ }\bibfield  {title} {\enquote {\bibinfo {title}
  {Band touching from real-space topology in frustrated hopping models},}\
  }\href {\doibase 10.1103/PhysRevB.78.125104} {\bibfield  {journal} {\bibinfo
  {journal} {Phys. Rev. B}\ }\textbf {\bibinfo {volume} {78}},\ \bibinfo
  {pages} {125104} (\bibinfo {year} {2008})}\BibitemShut {NoStop}%
\bibitem [{\citenamefont {Green}\ \emph {et~al.}(2010)\citenamefont {Green},
  \citenamefont {Santos},\ and\ \citenamefont {Chamon}}]{green2010isolated}%
  \BibitemOpen
  \bibfield  {author} {\bibinfo {author} {\bibfnamefont {Dmitry}\ \bibnamefont
  {Green}}, \bibinfo {author} {\bibfnamefont {Luiz}\ \bibnamefont {Santos}}, \
  and\ \bibinfo {author} {\bibfnamefont {Claudio}\ \bibnamefont {Chamon}},\
  }\bibfield  {title} {\enquote {\bibinfo {title} {Isolated flat bands and
  spin-1 conical bands in two-dimensional lattices},}\ }\href {\doibase
  10.1103/PhysRevB.82.075104} {\bibfield  {journal} {\bibinfo  {journal} {Phys.
  Rev. B}\ }\textbf {\bibinfo {volume} {82}},\ \bibinfo {pages} {075104}
  (\bibinfo {year} {2010})}\BibitemShut {NoStop}%
\bibitem [{\citenamefont {Mur-Petit}\ and\ \citenamefont
  {Molina}(2014)}]{murpetit2014chiral}%
  \BibitemOpen
  \bibfield  {author} {\bibinfo {author} {\bibfnamefont {Jordi}\ \bibnamefont
  {Mur-Petit}}\ and\ \bibinfo {author} {\bibfnamefont {Rafael~A.}\ \bibnamefont
  {Molina}},\ }\bibfield  {title} {\enquote {\bibinfo {title} {Chiral bound
  states in the continuum},}\ }\href {\doibase 10.1103/PhysRevB.90.035434}
  {\bibfield  {journal} {\bibinfo  {journal} {Phys. Rev. B}\ }\textbf {\bibinfo
  {volume} {90}},\ \bibinfo {pages} {035434} (\bibinfo {year}
  {2014})}\BibitemShut {NoStop}%
\bibitem [{\citenamefont {Poli}\ \emph {et~al.}(2017)\citenamefont {Poli},
  \citenamefont {Schomerus}, \citenamefont {Bellec}, \citenamefont {Kuhl},\
  and\ \citenamefont {Mortessagne}}]{poli2017partial}%
  \BibitemOpen
  \bibfield  {author} {\bibinfo {author} {\bibfnamefont {Charles}\ \bibnamefont
  {Poli}}, \bibinfo {author} {\bibfnamefont {Henning}\ \bibnamefont
  {Schomerus}}, \bibinfo {author} {\bibfnamefont {Matthieu}\ \bibnamefont
  {Bellec}}, \bibinfo {author} {\bibfnamefont {Ulrich}\ \bibnamefont {Kuhl}}, \
  and\ \bibinfo {author} {\bibfnamefont {Fabrice}\ \bibnamefont
  {Mortessagne}},\ }\bibfield  {title} {\enquote {\bibinfo {title} {Partial
  chiral symmetry-breaking as a route to spectrally isolated topological defect
  states in two-dimensional artificial materials},}\ }\href
  {http://stacks.iop.org/2053-1583/4/i=2/a=025008} {\bibfield  {journal}
  {\bibinfo  {journal} {2D Mat.}\ }\textbf {\bibinfo {volume} {4}},\ \bibinfo
  {pages} {025008} (\bibinfo {year} {2017})}\BibitemShut {NoStop}%
\bibitem [{\citenamefont {Leykam}\ \emph
  {et~al.}(2017{\natexlab{a}})\citenamefont {Leykam}, \citenamefont {Bodyfelt},
  \citenamefont {Desyatnikov},\ and\ \citenamefont
  {Flach}}]{leykam2017localization}%
  \BibitemOpen
  \bibfield  {author} {\bibinfo {author} {\bibfnamefont {Daniel}\ \bibnamefont
  {Leykam}}, \bibinfo {author} {\bibfnamefont {Joshua~D.}\ \bibnamefont
  {Bodyfelt}}, \bibinfo {author} {\bibfnamefont {Anton~S.}\ \bibnamefont
  {Desyatnikov}}, \ and\ \bibinfo {author} {\bibfnamefont {Sergej}\
  \bibnamefont {Flach}},\ }\bibfield  {title} {\enquote {\bibinfo {title}
  {Localization of weakly disordered flat band states},}\ }\href {\doibase
  10.1140/epjb/e2016-70551-2} {\bibfield  {journal} {\bibinfo  {journal} {Eur.
  Phys. J. B}\ }\textbf {\bibinfo {volume} {90}},\ \bibinfo {pages} {1}
  (\bibinfo {year} {2017}{\natexlab{a}})}\BibitemShut {NoStop}%
\bibitem [{\citenamefont {Read}(2017)}]{read2017compactly}%
  \BibitemOpen
  \bibfield  {author} {\bibinfo {author} {\bibfnamefont {N.}~\bibnamefont
  {Read}},\ }\bibfield  {title} {\enquote {\bibinfo {title} {Compactly
  supported wannier functions and algebraic k -theory},}\ }\href {\doibase
  10.1103/physrevb.95.115309} {\bibfield  {journal} {\bibinfo  {journal} {Phys.
  Rev. B}\ }\textbf {\bibinfo {volume} {95}} (\bibinfo {year} {2017}),\
  10.1103/physrevb.95.115309}\BibitemShut {NoStop}%
\bibitem [{\citenamefont {Lieb}(1989)}]{lieb1989two}%
  \BibitemOpen
  \bibfield  {author} {\bibinfo {author} {\bibfnamefont {Elliott~H.}\
  \bibnamefont {Lieb}},\ }\bibfield  {title} {\enquote {\bibinfo {title} {Two
  theorems on the hubbard model},}\ }\href {\doibase
  10.1103/PhysRevLett.62.1201} {\bibfield  {journal} {\bibinfo  {journal}
  {Phys. Rev. Lett.}\ }\textbf {\bibinfo {volume} {62}},\ \bibinfo {pages}
  {1201--1204} (\bibinfo {year} {1989})}\BibitemShut {NoStop}%
\bibitem [{\citenamefont {Griffiths}(1969)}]{griffiths1969nonanalytic}%
  \BibitemOpen
  \bibfield  {author} {\bibinfo {author} {\bibfnamefont {Robert~B.}\
  \bibnamefont {Griffiths}},\ }\bibfield  {title} {\enquote {\bibinfo {title}
  {Nonanalytic behavior above the critical point in a random ising
  ferromagnet},}\ }\href {\doibase 10.1103/PhysRevLett.23.17} {\bibfield
  {journal} {\bibinfo  {journal} {Phys. Rev. Lett.}\ }\textbf {\bibinfo
  {volume} {23}},\ \bibinfo {pages} {17--19} (\bibinfo {year}
  {1969})}\BibitemShut {NoStop}%
\bibitem [{\citenamefont {Vidal}\ \emph {et~al.}(1998)\citenamefont {Vidal},
  \citenamefont {Mosseri},\ and\ \citenamefont {Dou\ifmmode~\mbox{\c{c}}\else
  \c{c}\fi{}ot}}]{vidal1998aharonov}%
  \BibitemOpen
  \bibfield  {author} {\bibinfo {author} {\bibfnamefont {Julien}\ \bibnamefont
  {Vidal}}, \bibinfo {author} {\bibfnamefont {R\'emy}\ \bibnamefont {Mosseri}},
  \ and\ \bibinfo {author} {\bibfnamefont {Benoit}\ \bibnamefont
  {Dou\ifmmode~\mbox{\c{c}}\else \c{c}\fi{}ot}},\ }\bibfield  {title} {\enquote
  {\bibinfo {title} {Aharonov-bohm cages in two-dimensional structures},}\
  }\href {\doibase 10.1103/PhysRevLett.81.5888} {\bibfield  {journal} {\bibinfo
   {journal} {Phys. Rev. Lett.}\ }\textbf {\bibinfo {volume} {81}},\ \bibinfo
  {pages} {5888--5891} (\bibinfo {year} {1998})}\BibitemShut {NoStop}%
\bibitem [{\citenamefont {Vidal}\ \emph {et~al.}(2001)\citenamefont {Vidal},
  \citenamefont {Butaud}, \citenamefont {Dou\ifmmode~\mbox{\c{c}}\else
  \c{c}\fi{}ot},\ and\ \citenamefont {Mosseri}}]{vidal2001disorder}%
  \BibitemOpen
  \bibfield  {author} {\bibinfo {author} {\bibfnamefont {Julien}\ \bibnamefont
  {Vidal}}, \bibinfo {author} {\bibfnamefont {Patrick}\ \bibnamefont {Butaud}},
  \bibinfo {author} {\bibfnamefont {Benoit}\ \bibnamefont
  {Dou\ifmmode~\mbox{\c{c}}\else \c{c}\fi{}ot}}, \ and\ \bibinfo {author}
  {\bibfnamefont {R\'emy}\ \bibnamefont {Mosseri}},\ }\bibfield  {title}
  {\enquote {\bibinfo {title} {Disorder and interactions in aharonov-bohm
  cages},}\ }\href {\doibase 10.1103/PhysRevB.64.155306} {\bibfield  {journal}
  {\bibinfo  {journal} {Phys. Rev. B}\ }\textbf {\bibinfo {volume} {64}},\
  \bibinfo {pages} {155306} (\bibinfo {year} {2001})}\BibitemShut {NoStop}%
\bibitem [{\citenamefont {Lan}\ \emph {et~al.}(2012)\citenamefont {Lan},
  \citenamefont {Goldman},\ and\ \citenamefont
  {\"Ohberg}}]{lan2012coexistence}%
  \BibitemOpen
  \bibfield  {author} {\bibinfo {author} {\bibfnamefont {Zhihao}\ \bibnamefont
  {Lan}}, \bibinfo {author} {\bibfnamefont {Nathan}\ \bibnamefont {Goldman}}, \
  and\ \bibinfo {author} {\bibfnamefont {Patrik}\ \bibnamefont {\"Ohberg}},\
  }\bibfield  {title} {\enquote {\bibinfo {title} {Coexistence of
  spin-$\frac{1}{2}$ and spin-1 dirac-weyl fermions in the edge-centered
  honeycomb lattice},}\ }\href {\doibase 10.1103/PhysRevB.85.155451} {\bibfield
   {journal} {\bibinfo  {journal} {Phys. Rev. B}\ }\textbf {\bibinfo {volume}
  {85}},\ \bibinfo {pages} {155451} (\bibinfo {year} {2012})}\BibitemShut
  {NoStop}%
\bibitem [{\citenamefont {Lin}\ \emph {et~al.}(2009{\natexlab{a}})\citenamefont
  {Lin}, \citenamefont {Compton}, \citenamefont {Perry}, \citenamefont
  {Phillips}, \citenamefont {Porto},\ and\ \citenamefont
  {Spielman}}]{lin2009bose}%
  \BibitemOpen
  \bibfield  {author} {\bibinfo {author} {\bibfnamefont {Y.-J.}\ \bibnamefont
  {Lin}}, \bibinfo {author} {\bibfnamefont {R.~L.}\ \bibnamefont {Compton}},
  \bibinfo {author} {\bibfnamefont {A.~R.}\ \bibnamefont {Perry}}, \bibinfo
  {author} {\bibfnamefont {W.~D.}\ \bibnamefont {Phillips}}, \bibinfo {author}
  {\bibfnamefont {J.~V.}\ \bibnamefont {Porto}}, \ and\ \bibinfo {author}
  {\bibfnamefont {I.~B.}\ \bibnamefont {Spielman}},\ }\bibfield  {title}
  {\enquote {\bibinfo {title} {Bose-einstein condensate in a uniform
  light-induced vector potential},}\ }\href {\doibase
  10.1103/PhysRevLett.102.130401} {\bibfield  {journal} {\bibinfo  {journal}
  {Phys. Rev. Lett.}\ }\textbf {\bibinfo {volume} {102}},\ \bibinfo {pages}
  {130401} (\bibinfo {year} {2009}{\natexlab{a}})}\BibitemShut {NoStop}%
\bibitem [{\citenamefont {Lin}\ \emph {et~al.}(2009{\natexlab{b}})\citenamefont
  {Lin}, \citenamefont {Compton}, \citenamefont {Jimenez-Garcia}, \citenamefont
  {Porto},\ and\ \citenamefont {Spielman}}]{lin2009synthetic}%
  \BibitemOpen
  \bibfield  {author} {\bibinfo {author} {\bibfnamefont {Y.~J.}\ \bibnamefont
  {Lin}}, \bibinfo {author} {\bibfnamefont {R.~L.}\ \bibnamefont {Compton}},
  \bibinfo {author} {\bibfnamefont {K.}~\bibnamefont {Jimenez-Garcia}},
  \bibinfo {author} {\bibfnamefont {J.~V.}\ \bibnamefont {Porto}}, \ and\
  \bibinfo {author} {\bibfnamefont {I.~B.}\ \bibnamefont {Spielman}},\
  }\bibfield  {title} {\enquote {\bibinfo {title} {Synthetic magnetic fields
  for ultracold neutral atoms},}\ }\href
  {http://dx.doi.org/10.1038/nature08609} {\bibfield  {journal} {\bibinfo
  {journal} {Nature}\ }\textbf {\bibinfo {volume} {462}},\ \bibinfo {pages}
  {628--632} (\bibinfo {year} {2009}{\natexlab{b}})}\BibitemShut {NoStop}%
\bibitem [{\citenamefont {Aidelsburger}\ \emph {et~al.}(2013)\citenamefont
  {Aidelsburger}, \citenamefont {Atala}, \citenamefont {Lohse}, \citenamefont
  {Barreiro}, \citenamefont {Paredes},\ and\ \citenamefont
  {Bloch}}]{aidelsburger2013realization}%
  \BibitemOpen
  \bibfield  {author} {\bibinfo {author} {\bibfnamefont {M.}~\bibnamefont
  {Aidelsburger}}, \bibinfo {author} {\bibfnamefont {M.}~\bibnamefont {Atala}},
  \bibinfo {author} {\bibfnamefont {M.}~\bibnamefont {Lohse}}, \bibinfo
  {author} {\bibfnamefont {J.~T.}\ \bibnamefont {Barreiro}}, \bibinfo {author}
  {\bibfnamefont {B.}~\bibnamefont {Paredes}}, \ and\ \bibinfo {author}
  {\bibfnamefont {I.}~\bibnamefont {Bloch}},\ }\bibfield  {title} {\enquote
  {\bibinfo {title} {Realization of the hofstadter hamiltonian with ultracold
  atoms in optical lattices},}\ }\href {\doibase
  10.1103/PhysRevLett.111.185301} {\bibfield  {journal} {\bibinfo  {journal}
  {Phys. Rev. Lett.}\ }\textbf {\bibinfo {volume} {111}},\ \bibinfo {pages}
  {185301} (\bibinfo {year} {2013})}\BibitemShut {NoStop}%
\bibitem [{\citenamefont {Miyake}\ \emph {et~al.}(2013)\citenamefont {Miyake},
  \citenamefont {Siviloglou}, \citenamefont {Kennedy}, \citenamefont {Burton},\
  and\ \citenamefont {Ketterle}}]{miyake2013realizing}%
  \BibitemOpen
  \bibfield  {author} {\bibinfo {author} {\bibfnamefont {Hirokazu}\
  \bibnamefont {Miyake}}, \bibinfo {author} {\bibfnamefont {Georgios~A.}\
  \bibnamefont {Siviloglou}}, \bibinfo {author} {\bibfnamefont {Colin~J.}\
  \bibnamefont {Kennedy}}, \bibinfo {author} {\bibfnamefont {William~Cody}\
  \bibnamefont {Burton}}, \ and\ \bibinfo {author} {\bibfnamefont {Wolfgang}\
  \bibnamefont {Ketterle}},\ }\bibfield  {title} {\enquote {\bibinfo {title}
  {Realizing the harper hamiltonian with laser-assisted tunneling in optical
  lattices},}\ }\href {\doibase 10.1103/PhysRevLett.111.185302} {\bibfield
  {journal} {\bibinfo  {journal} {Phys. Rev. Lett.}\ }\textbf {\bibinfo
  {volume} {111}},\ \bibinfo {pages} {185302} (\bibinfo {year}
  {2013})}\BibitemShut {NoStop}%
\bibitem [{\citenamefont {Aidelsburger}\ \emph {et~al.}(2015)\citenamefont
  {Aidelsburger}, \citenamefont {Lohse}, \citenamefont {Schweizer},
  \citenamefont {Atala}, \citenamefont {Barreiro}, \citenamefont {Nascimbene},
  \citenamefont {Cooper}, \citenamefont {Bloch},\ and\ \citenamefont
  {Goldman}}]{aidelsburger2015measuring}%
  \BibitemOpen
  \bibfield  {author} {\bibinfo {author} {\bibfnamefont {M.}~\bibnamefont
  {Aidelsburger}}, \bibinfo {author} {\bibfnamefont {M.}~\bibnamefont {Lohse}},
  \bibinfo {author} {\bibfnamefont {C.}~\bibnamefont {Schweizer}}, \bibinfo
  {author} {\bibfnamefont {M.}~\bibnamefont {Atala}}, \bibinfo {author}
  {\bibfnamefont {J.~T.}\ \bibnamefont {Barreiro}}, \bibinfo {author}
  {\bibfnamefont {S.}~\bibnamefont {Nascimbene}}, \bibinfo {author}
  {\bibfnamefont {N.~R.}\ \bibnamefont {Cooper}}, \bibinfo {author}
  {\bibfnamefont {I.}~\bibnamefont {Bloch}}, \ and\ \bibinfo {author}
  {\bibfnamefont {N.}~\bibnamefont {Goldman}},\ }\bibfield  {title} {\enquote
  {\bibinfo {title} {Measuring the chern number of hofstadter bands with
  ultracold bosonic atoms},}\ }\href {http://dx.doi.org/10.1038/nphys3171}
  {\bibfield  {journal} {\bibinfo  {journal} {Nat. Phys.}\ }\textbf {\bibinfo
  {volume} {11}},\ \bibinfo {pages} {162--166} (\bibinfo {year}
  {2015})}\BibitemShut {NoStop}%
\bibitem [{\citenamefont {Bloch}\ \emph {et~al.}(2012)\citenamefont {Bloch},
  \citenamefont {Dalibard},\ and\ \citenamefont
  {Nascimbene}}]{bloch2012quantum}%
  \BibitemOpen
  \bibfield  {author} {\bibinfo {author} {\bibfnamefont {Immanuel}\
  \bibnamefont {Bloch}}, \bibinfo {author} {\bibfnamefont {Jean}\ \bibnamefont
  {Dalibard}}, \ and\ \bibinfo {author} {\bibfnamefont {Sylvain}\ \bibnamefont
  {Nascimbene}},\ }\bibfield  {title} {\enquote {\bibinfo {title} {Quantum
  simulations with ultracold quantum gases},}\ }\href
  {http://dx.doi.org/10.1038/nphys2259} {\bibfield  {journal} {\bibinfo
  {journal} {Nat. Phys.}\ }\textbf {\bibinfo {volume} {8}},\ \bibinfo {pages}
  {267--276} (\bibinfo {year} {2012})}\BibitemShut {NoStop}%
\bibitem [{\citenamefont {Struck}\ \emph {et~al.}(2012)\citenamefont {Struck},
  \citenamefont {\"Olschl\"ager}, \citenamefont {Weinberg}, \citenamefont
  {Hauke}, \citenamefont {Simonet}, \citenamefont {Eckardt}, \citenamefont
  {Lewenstein}, \citenamefont {Sengstock},\ and\ \citenamefont
  {Windpassinger}}]{struck2012tunable}%
  \BibitemOpen
  \bibfield  {author} {\bibinfo {author} {\bibfnamefont {J.}~\bibnamefont
  {Struck}}, \bibinfo {author} {\bibfnamefont {C.}~\bibnamefont
  {\"Olschl\"ager}}, \bibinfo {author} {\bibfnamefont {M.}~\bibnamefont
  {Weinberg}}, \bibinfo {author} {\bibfnamefont {P.}~\bibnamefont {Hauke}},
  \bibinfo {author} {\bibfnamefont {J.}~\bibnamefont {Simonet}}, \bibinfo
  {author} {\bibfnamefont {A.}~\bibnamefont {Eckardt}}, \bibinfo {author}
  {\bibfnamefont {M.}~\bibnamefont {Lewenstein}}, \bibinfo {author}
  {\bibfnamefont {K.}~\bibnamefont {Sengstock}}, \ and\ \bibinfo {author}
  {\bibfnamefont {P.}~\bibnamefont {Windpassinger}},\ }\bibfield  {title}
  {\enquote {\bibinfo {title} {Tunable gauge potential for neutral and spinless
  particles in driven optical lattices},}\ }\href {\doibase
  10.1103/PhysRevLett.108.225304} {\bibfield  {journal} {\bibinfo  {journal}
  {Phys. Rev. Lett.}\ }\textbf {\bibinfo {volume} {108}},\ \bibinfo {pages}
  {225304} (\bibinfo {year} {2012})}\BibitemShut {NoStop}%
\bibitem [{\citenamefont {Celi}\ \emph {et~al.}(2014)\citenamefont {Celi},
  \citenamefont {Massignan}, \citenamefont {Ruseckas}, \citenamefont {Goldman},
  \citenamefont {Spielman}, \citenamefont {Juzeli\ifmmode~\bar{u}\else
  \={u}\fi{}nas},\ and\ \citenamefont {Lewenstein}}]{celi2014synthetic}%
  \BibitemOpen
  \bibfield  {author} {\bibinfo {author} {\bibfnamefont {A.}~\bibnamefont
  {Celi}}, \bibinfo {author} {\bibfnamefont {P.}~\bibnamefont {Massignan}},
  \bibinfo {author} {\bibfnamefont {J.}~\bibnamefont {Ruseckas}}, \bibinfo
  {author} {\bibfnamefont {N.}~\bibnamefont {Goldman}}, \bibinfo {author}
  {\bibfnamefont {I.~B.}\ \bibnamefont {Spielman}}, \bibinfo {author}
  {\bibfnamefont {G.}~\bibnamefont {Juzeli\ifmmode~\bar{u}\else
  \={u}\fi{}nas}}, \ and\ \bibinfo {author} {\bibfnamefont {M.}~\bibnamefont
  {Lewenstein}},\ }\bibfield  {title} {\enquote {\bibinfo {title} {Synthetic
  gauge fields in synthetic dimensions},}\ }\href {\doibase
  10.1103/PhysRevLett.112.043001} {\bibfield  {journal} {\bibinfo  {journal}
  {Phys. Rev. Lett.}\ }\textbf {\bibinfo {volume} {112}},\ \bibinfo {pages}
  {043001} (\bibinfo {year} {2014})}\BibitemShut {NoStop}%
\bibitem [{\citenamefont {Golshani}\ \emph {et~al.}(2014)\citenamefont
  {Golshani}, \citenamefont {Weimann}, \citenamefont {Jafari}, \citenamefont
  {Nezhad}, \citenamefont {Langari}, \citenamefont {Bahrampour}, \citenamefont
  {Eichelkraut}, \citenamefont {Mahdavi},\ and\ \citenamefont
  {Szameit}}]{golshani2014impact}%
  \BibitemOpen
  \bibfield  {author} {\bibinfo {author} {\bibfnamefont {M.}~\bibnamefont
  {Golshani}}, \bibinfo {author} {\bibfnamefont {S.}~\bibnamefont {Weimann}},
  \bibinfo {author} {\bibfnamefont {Kh.}\ \bibnamefont {Jafari}}, \bibinfo
  {author} {\bibfnamefont {M.~Khazaei}\ \bibnamefont {Nezhad}}, \bibinfo
  {author} {\bibfnamefont {A.}~\bibnamefont {Langari}}, \bibinfo {author}
  {\bibfnamefont {A.~R.}\ \bibnamefont {Bahrampour}}, \bibinfo {author}
  {\bibfnamefont {T.}~\bibnamefont {Eichelkraut}}, \bibinfo {author}
  {\bibfnamefont {S.~M.}\ \bibnamefont {Mahdavi}}, \ and\ \bibinfo {author}
  {\bibfnamefont {A.}~\bibnamefont {Szameit}},\ }\bibfield  {title} {\enquote
  {\bibinfo {title} {Impact of loss on the wave dynamics in photonic waveguide
  lattices},}\ }\href {\doibase 10.1103/PhysRevLett.113.123903} {\bibfield
  {journal} {\bibinfo  {journal} {Phys. Rev. Lett.}\ }\textbf {\bibinfo
  {volume} {113}},\ \bibinfo {pages} {123903} (\bibinfo {year}
  {2014})}\BibitemShut {NoStop}%
\bibitem [{\citenamefont {Longhi}(2013)}]{longhi2013effective}%
  \BibitemOpen
  \bibfield  {author} {\bibinfo {author} {\bibfnamefont {Stefano}\ \bibnamefont
  {Longhi}},\ }\bibfield  {title} {\enquote {\bibinfo {title} {Effective
  magnetic fields for photons in waveguide and coupled resonator lattices},}\
  }\href {\doibase 10.1364/OL.38.003570} {\bibfield  {journal} {\bibinfo
  {journal} {Opt. Lett.}\ }\textbf {\bibinfo {volume} {38}},\ \bibinfo {pages}
  {3570--3573} (\bibinfo {year} {2013})}\BibitemShut {NoStop}%
\bibitem [{\citenamefont {Khomeriki}\ and\ \citenamefont
  {Flach}(2016)}]{khomeriki2016landau}%
  \BibitemOpen
  \bibfield  {author} {\bibinfo {author} {\bibfnamefont {Ramaz}\ \bibnamefont
  {Khomeriki}}\ and\ \bibinfo {author} {\bibfnamefont {Sergej}\ \bibnamefont
  {Flach}},\ }\bibfield  {title} {\enquote {\bibinfo {title} {Landau-zener
  bloch oscillations with perturbed flat bands},}\ }\href {\doibase
  10.1103/PhysRevLett.116.245301} {\bibfield  {journal} {\bibinfo  {journal}
  {Phys. Rev. Lett.}\ }\textbf {\bibinfo {volume} {116}},\ \bibinfo {pages}
  {245301} (\bibinfo {year} {2016})}\BibitemShut {NoStop}%
\bibitem [{\citenamefont {Leykam}\ \emph
  {et~al.}(2017{\natexlab{b}})\citenamefont {Leykam}, \citenamefont {Flach},\
  and\ \citenamefont {Chong}}]{leykam2017flat}%
  \BibitemOpen
  \bibfield  {author} {\bibinfo {author} {\bibfnamefont {Daniel}\ \bibnamefont
  {Leykam}}, \bibinfo {author} {\bibfnamefont {Sergej}\ \bibnamefont {Flach}},
  \ and\ \bibinfo {author} {\bibfnamefont {Y.~D.}\ \bibnamefont {Chong}},\
  }\href {https://arxiv.org/abs/1704.00896} {\enquote {\bibinfo {title} {Flat
  bands in lattices with non-hermitian coupling},}\ } (\bibinfo {year}
  {2017}{\natexlab{b}}),\ \Eprint {http://arxiv.org/abs/1704.00896}
  {arXiv:1704.00896 [physics.optics]} \BibitemShut {NoStop}%
\end{thebibliography}%

\end{document}